\documentclass[12pt,a4paper]{article}
\usepackage{latexsym,amssymb,amsmath,amsthm}

\newtheorem{theorem}{Theorem}

\renewcommand{\proof}{\noindent {\bf Proof. }}

\begin{document}
\title{On the complexity of computing prime tables on a Turing machine\footnote{Translated
 from the Russian original published in: Prikladnaya diskretnaya matematika (Applied Discrete Mathematics). 2016. N.\,1(31). 86--91.}}
\date{}
\author{Igor S. Sergeev\footnote{e-mail: isserg@gmail.com}}
%\thanks{}
\maketitle

\begin{abstract}
We prove that the complexity of computing the table of primes
between $1$ and $n$ on a multitape Turing machine is $O(n\log n)$.
\end{abstract}

\section*{Introduction}

Computing of a table of prime numbers is a necessary preliminary
stage in modern factorization algorithms (see~\cite{cp}); among
other applications we can indicate the factorial calculation
(see~\cite[\S6.5]{sgv}).

The log-RAM computational model (that is, a RAM-program executing
arithmetic operations with $O(\log n)$-size operands in time
$O(1)$, where $n$ is an input size, see~\cite{ahu,fu}) is commonly
used for theoretical complexity bounds due to a closeness to
standard computer computations. The best known upper complexity
bound of the table of primes up to $n$ in the log-RAM model is
$O(n/\log\log n)$~\cite{pr} (see also~\cite[\S 3.2]{cp}).

The multitape Turing machine (MTM) model serves as a universal
mean for analysis of computational algorithms. The definition of
MTM see, e.g., in~\cite{ahu,sgv}. Informally speaking, MTM
contains several (i.e. $O(1)$) potentially infinite tapes of
binary symbols accessed by pointers and a processor executing
elementary instructions. The processor uses tapes as a memory. A
set of instructions includes: shift of a pointer by one position
to the left or to the right, read of a symbol from a pointed
position, writing a symbol to a pointed position, a binary boolean
operation over two symbols, a conditional operation (selection of
instruction to execute depending on the value of a
symbol)\footnote{Really, the functionality of the Turing machine
as an abstract model of computations is much wider. The given
description appeals to known implementations of MTM, see,
e.g.~\cite{sgv}.}. Complexity is measured by the number of
instructions.

Sch\"onhage~\cite{sgv} adapted the Eratostene sieve to estimate
the complexity $P(n)$ of computing the table of primes up to $n$
as $O(n\log^2n\log\log n)$. Though, he made a remark that this
bound may be improved via a more accurate analysis of the
algorithm\footnote{Since the aim of~\cite{sgv} was to compute $n!$
with $O(M(n\log n))$ complexity, the mentioned bound for $P(n)$
suited until the complexity $M(n)$ of multiplication of $n$-bit
integers was proved to be $o(n\log n \log\log n)$.}.

The authors of~\cite{ft} adapted to MTM the method~\cite{ab} based
on a special ``quadratic'' sieve and proved the bound
$P(n)=O(n\log^2n/\log\log n)$.

Actually, an accurate analysis allows to estimate the complexity
of the method~\cite{sgv} as $O(n\log^2 n)$. This is the same bound
as for a simple version of the method~\cite{ft}. Yet, a trick of
avoiding repeated processing of smooth numbers due to
Pritchard~\cite{pr} (see also~\cite[\S 3.2]{cp}) allows to further
reduce the complexity to $O(n\log^2 n/ \log\log n)$, as in the
method~\cite{ft}.

The paper~\cite{ft} proposes one more method providing the
complexity bound $P(n)=O(M(n\log n))$, where $M(n)$ is the
complexity of multiplication of $n$-bit integers. This method
receives a priority under assumption $M(n)=o(n\log n)$. However,
today the plausibility of such hypothesis seems to be unlikely.

Here we prove the bound $P(n)=O(n\log n)$. Then, it follows that
the complexity of computing the table of composite numbers between
1 and $n$ is $\Theta(n\log n)$. One more consequence: computing of
primes does not affect the order of complexity of $n!$, which is
obviously (from the size of the problem) at least $n\log n$.

\section*{General scheme of computations}

Due to exploiting of the Eratostene sieve or the sieve~\cite{ab}
methods~\cite{sgv,ft} lead to overestimated complexity bounds, as
a result of multiple reuse of composite numbers. Hence, it may be
profitable to base on an irredundant way of enumerating composite
numbers proposed by Mairson~\cite{ma}. This strategy was shown to
be efficient in~\cite{pr} (see also~\cite[\S 3.2]{cp}).

Let $p_i$ denote subsequent prime numbers: $p_1=2$, $p_2=3$, etc.
A procedure due to Mairson lists all composite numbers in a given
interval avoiding repetitions. Namely, for each $i$ it produces a
chain $C_i$ of numbers $C_{i,j} = k_j \cdot p_i$ not dividing by
primes smaller than $p_i$. Factors $k_j$ may be determined from
the results of preceding stages of the sieving by condition $k_j
\notin \bigcup_{i'<i} C_{i'}$.

Essentially, the following algorithm is an adaptation of the
algorithm from~\cite[\S 3.2]{cp} based on ideas~\cite{ma,pr}.

In $n$ cells of some tape we store indicators of primality of
numbers from 1 to~$n$. Initially, the tape is filled by zeros.
After completion of the algorithm, 0 in position $k$ indicates
that $k$ is prime, and 1 implies that $k$ is composite.

Subsequently incrementing index $i$ we scan the tape putting ones
in positions $C_{i,j}$. Since the whole tape scanning requires
$O(n)$ instructions, we gradually reduce the length of initial
intervals of the tape to scan while $i$ grows. Nevertheless, we
still succeed in constructing a set of multiples $k_j$ for a
subsequent step.

After some chain (which is a sorted list) $C_i$ is computed, it is
copied to another tape of the MTM. Next, we implement a standard
process of pairwise mergings of the chains into sorted lists until
the total number of chains is less than $\log n$. To reduce the
cost of comparisons in the sorting, we store and process the lists
in compressed form. Finally, each obtained chain is to be mapped
to the tape of primality indicators.

\section*{Main result}

Below, we use some standard facts concerning the distribution of
prime numbers, see, e.g.~\cite{rs,hand}. As usual, $\pi(n)$
denotes the number of primes smaller of equal to $n$.

\begin{theorem}
The complexity of computing the table of primes from $1$ to $n$ is
$O(n\log n)$.
\end{theorem}

\proof Let us make a few preliminary remarks.

While a pointer moves along a tape, one has to update the index of
its position. Thus, recall that incrementing or decrementing of a
number may be executed via $O(1)$ bit operations, on average.
Indeed, half of the cases requires updating just the lowest bit,
$1/4$ of the cases requires two bits to be updated, $1/8$ of the
cases requires updating of three bits, etc.

If a pointer should be moved to a prescribed position, one has to
permanently compare indices of the current and the target
positions. This comparison also may be implemented with complexity
$O(1)$: in half of the cases the lowest bits should be compared,
in $1/4$ of the cases two bit comparisons should be done, etc.

Now, we are prepared to state the algorithm.

0) The general purpose is to eventually form on the type $T$ 1-bit
indicators $a(k)$ of primality of numbers $k$. Initially, the type
is filled as $a(2)=\ldots=a(n)=0$.

1) For any prime number $p_i \le \log n$ we scan the tape $T$ and
set $a(kp_i)=1$, $k>1$. The number $p_i$ may be determined as a
position of $i$-th zero symbol. While scanning the tape we update
the index of position modulo $p_i$, that is, increment the current
index and compare it with $p_i$. If the equality holds, then write
1 into the corresponding cell and zero the position modular index.
The complexity of this stage is $O(n\cdot\pi(\log n)) = O(n\log n/
\log\log n)$.

2) For any larger prime number $\log n < p_i \le \sqrt n$ we scan
only first $n/p_i$ positions of the tape. As before, the index of
$i$-th zero position is $p_i$ itself. Moving further along the
tape we perform two actions. First, we write to another tape a
chain $C_i$ of composite numbers $kp_i$ not dividing by primes
less than $p_i$. The condition $a(k)=0$ for $k \ge p_i$ serves as
a criterion of being an element of $C_i$. Second, we write ones in
positions with indices divisible by $p_i$.

In representation of chains $C_i$ we use compression. Consider the
following partition of primes into groups: $j$-th group is
constituted by primes $p \in [2^j,\, 2^{j+1})$. Then, we partition
a chain referred to a prime from $j$-th group into $n2^{-2j}$
sections (up to rounding\footnote{Here and further we omit
roundings in formulae (if it doesn't affect results) to make
presentation cleaner.}): each section contains numbers differing
only by $2j$ lowest bits. Other $\log n-2j$ bits may be determined
by the index of a section. Thus, to write a new number into a
chain we practically write its $2j$ lowest bits into appropriate
section.\footnote{Technically, one can use separating bits: say,
separate sections by ones and separate numbers inside a section by
zeros.}

A chain $C_i$ contains at most\footnote{Symbol $\asymp$ denotes
equal orders of growth.}
$$ \frac{n}{p_i} \prod_{j=1}^{i-1} \left( 1 - \frac1{p_j} \right) \asymp \frac{n}{p_i \log p_i}  $$
numbers by order of magnitude (the equality holds due to a
prominent result by Chebyshov and Mertens). The computation
proceeds as follows. While scanning the tape $T$ we compute
distances between nearest zeros. When we achieve a cell with zero
the accumulated distance $s$ must be multiplied by $p_i$. Such
multiplication may be trivially done by a ``shift and add'' method
via $O(\log p_i \cdot \log s)$ operations, if the length of $s$ is
known (it is convenient to store this length on another tape).
Recall that if a sum of $t$ numbers is at most $th$, then the sum
of their logarithms is at most $t\log h$ (the latter sum is
maximal when all summands are equal). Therefore, to bound from
above the complexity of multiplications it suffices to replace $s$
by a mean value of distance which is of order $\log p_i$. Hence,
the complexity is bounded by $O(n\log\log p_i/p_i)$.

At last, the complexity of creating of a new element of a chain,
that is, addition of the computed difference to the preceding
element and insertion of the new element into the chain, may be
(roughly) estimated as $O(\log n)$.

By summation over all indices $i$, the order of complexity of the
present stage may be upper estimated as
$$ n\sum_{i=\pi(\log n)+1}^{\pi(\sqrt n)} \frac{\log \log p_i}{p_i} + n\log n \sum_{i=\pi(\log n)+1}^{\pi(\sqrt n)} \frac{1}{p_i \log p_i} \asymp n\log n / \log\log n.  $$

3) After completion of the previous stage, we have about
$\pi(\sqrt n)$ sorted lists of composite numbers including totally
$$ n \sum_{i=\pi(\log n)+1}^{\pi(\sqrt n)} \frac{1}{p_i \log p_i} \asymp n/ \log\log n $$
elements, by order of magnitude. These lists are arranged in
approximately $(1/2)\log n$ groups (see above)~--- all lists in a
group obey the same formula of compression. Let us sort the
elements inside each group. By construction, $j$-th group contains
approximately $2^j/j$ chains and
$$ n \sum_{i=\pi(2^j)+1}^{\pi(2^{j+1})} \frac{1}{p_i \log p_i} \asymp n/j^2 $$
numbers, by order of magnitude. A merging tree (see,
e.g.~\cite[\S5.4]{kn}) allows to sort these chains via $O(n/j)$
comparison and rewriting operations, each operation is equivalent
to $O(j)$ bit operations.\footnote{Merging of $s$ sorted lists
with total number $N$ elements may be implemented via $O(N\log s)$
comparison and read/write operations~--- three tapes of MTM are
sufficient for this, more details see in~\cite{kn,sgv}.}

The total complexity of the stage is $O(n\log n)$, that is, $O(n)$
per group. (Note, that the compression reduces the complexity of
comparison in $j$-th group from $O(\log n)$ to $O(j)$ and the
order of complexity of the stage from $n\log n \log\log n$ to
$n\log n$.)

4) For any list $S$ obtained at the end of the preceding stage we
update the tape $T$ by setting $a(k)=1$ for $k \in S$. This
procedure may be performed via single browsing of the list and
single passage along the tape with $O(n)$ complexity. Thus, the
complexity of the current stage is $O(n\log n)$.

5) After completion of the previous stage the tape $T$ is fully
prepared: it has zeros exactly in positions with prime indices.
Now, one passage along the tape suffices to compose the table of
primes with complexity of order $\pi(n)\cdot \log n + n \asymp n$.
\qed
%\end{proof}

Research supported in part by RFBR, grant 14--01--00671a.


\begin{thebibliography}{99}

\bibitem{ahu}
Aho\;A., Hopkroft\;J., Ullman\;J. The design and analysis of
computer algorithms. Reading, MA: Addison-Wesley, 1974.

\bibitem{ab}
Atkin\;A.\,O.\,L., Bernstein\;D.\,J. Prime sieves using binary
quadratic forms~// Math. Comput. 2004. V.\,73(246).
P.\,1023--1030.

\bibitem{cp}
Crandall\;R., Pomerance\;C. Prime numbers: a computational
perspective. NY: Springer, 2005.

\bibitem{ft}
Farach-Colton\;M., Tsai\;M.-T. On the complexity of computing
prime tables~// Proc. ISAAC 2015. LNCS. V.\,9472. Berlin,
Heidelberg: Springer, 2015, P.\,677--688. {\sf arXiv:1504.05240.}

\bibitem{fu}
F\"urer\;M. How fast can we multiply large integers on an actual
computer?~// Proc. LATIN 2014. Theor. Comput. Sci. and Gen. Iss.
V.\,8392. Berlin, Heidelberg: Springer, 2014, P.\,660--670. {\sf
arXiv:1402.1811.}

\bibitem{kn}
Knuth\;D. The art of computer programming. Vol. 3. Sorting and
searching. Reading, MA: Addison-Wesley, 1998.

\bibitem{ma}
Mairson\;H.\,G. Some new upper bounds on the generation of prime
numbers~// Comm. ACM. 1977. V.\,20(9). P.\,664--669.

\bibitem{pr}
Pritchard\;P. A sublinear additive sieve for finding prime
numbers~// Comm. ACM. 1981. V.\,24(1). P.\,18--23.

\bibitem{rs}
Rosser\;J., Schoenfeld\;L. Approximate formulas for some functions
of prime numbers~// Ill. J. Math. 1962. V.\,6. P.\,64--94.

\bibitem{hand}
S\'andor\;J., Mitrinovi\'c\;D.\,S., Crstici\;B. Handbook of number
theory. I. Dordrecht: Springer, 2006.

\bibitem{sgv}
Sch\"onhage\;A., Grotefeld\;A.\,F.\,W., Vetter\;E. Fast
algorithms: a multitape Turing machine implementation. Mannheim,
Leipzig, Wien, Z\"urich: BI-Wissenschaftsverlag, 1994.

\end{thebibliography}
\end{document}